\documentclass[aps,prr,twocolumn,superscriptaddress,nofootinbib]{revtex4-1}
\usepackage{amsmath}
\usepackage{amssymb}
\usepackage{graphicx}
\usepackage{mathrsfs}
\usepackage{ntheorem}
\usepackage{mathtools}
\usepackage{enumerate}
\usepackage{enumitem}
\usepackage{times,txfonts}
\usepackage{subfigure}
\usepackage{upgreek}
\usepackage{tikz}
\usepackage[colorlinks,bookmarksopen,bookmarksnumbered,citecolor=blue, linkcolor=blue, urlcolor=blue]{hyperref}
\newcommand{\ket}[1]{|#1\rangle}
\newcommand{\bra}[1]{\langle #1|}
\newcommand{\Tr}{\mathrm{Tr}}

\newcommand{\abs}[1]{\lvert #1\rvert}

\def\CC{{\rm\kern.24em \vrule width.04em height1.46ex depth-.07ex \kern-.30em C}}
\def\RR{{\rm\kern.24em \vrule width.04em height1.46ex depth-.07ex
\kern-.30em R}}
\def\P{{\rm I\kern-.25em P}}

\begin{document}
\title{Optimal probabilistic distillation of quantum coherence}

\author{C. L. Liu}
\affiliation{Graduate School of China Academy of Engineering Physics, Beijing 100193, China}
\author{C. P. Sun}
\email{suncp@gscaep.ac.cn}
\affiliation{Graduate School of China Academy of Engineering Physics, Beijing 100193, China}
\affiliation{Beijing Computational Science Research Center, Beijing 100193, China}
\date{\today}
\begin{abstract}
We complete the task of optimal probabilistic coherence distillation protocol, whose aim is to transform a general state into a set of $n$-level maximally coherent states via strictly incoherent operations (SIO). Specifically, we present the necessary and sufficient condition for the transformation from a mixed state into a pure-state ensemble via SIO. With the aid of this condition and the simplex algorithm, we can accomplish the probabilistic distillation protocol by presenting an analytical expression of the maximal average distillable coherence of a general state and constructing the corresponding operation achieving this bound. Our protocol is a universal protocol since it can be applied to any coherence measure.
\end{abstract}
\maketitle

\section{Introduction}
Quantum coherence is an important feature of quantum mechanics which is responsible for the departure between the classical and quantum world. It is an essential component in quantum information processing \cite{Nielsen}, and plays a central role in various fields, such as quantum computation \cite{Shor,Grover}, quantum cryptography \cite{Bennett},  quantum metrology \cite{Giovannetti,Giovannetti1}, and quantum biology \cite{Lambert}. Recently, the resource theory of coherence has attracted a growing interest due to the rapid development of quantum information science \cite{Aberg1,Baumgratz,Streltsov}. The resource theory of coherence not only establishes a rigorous framework to quantify coherence but also provides a platform to understand  quantum coherence from a different perspective.

Any quantum resource theory is characterized by two fundamental ingredients, namely, free states and free operations \cite{Horodecki,Brandao,Chitambar}. For the resource theory of coherence, the free states are quantum states which are diagonal in a prefixed reference basis. The free operations are not uniquely specified. Motivated by suitable practical considerations, several free operations were presented \cite{Levi,Aberg1,Baumgratz,Vicente,Chitambar2,Chitambar3,Winter,Yadin,Marvian}. In this paper, we focus our attention on the strictly incoherent operations (SIO), which were first proposed in Ref. \cite{Winter}; in Ref. \cite{Yadin}, it has been shown that these operations neither create nor use coherence and have a physical interpretation in terms of interferometry. Thus the set of SIO is a physically well-motivated set of free operations for the resource theory of coherence.

In resource theories, much effort has been devoted to studying the distillation protocols. The distillation process is the process that extracts pure resource states from a general state via free operations. For the resource theory of coherence, various coherence distillation protocols were proposed. These protocols can be divided into two classes: The asymptotic coherence distillation and the one-shot coherence distillation. The asymptotic coherence distillation of pure states and mixed states by using free operations was studied in Refs. \cite{Yuan,Winter,Lami,Lami1,Zhao}. To relax several unreasonable assumptions of the asymptotic regime, i.e., unbounded number copies of identical states and collective operations, several one-shot coherence distillation protocols were proposed and explored \cite{Liu1,Liu2,Liu3,Fang,Regula,Regula1,Chitambar1,Torun,Regula2,Chen,Zhang,Du,Bu,Ziwen,Zhu,Du1}.

Here we study the optimal probabilistic coherence distillation, which is one of the one-shot coherence distillation protocols. The aim of this protocol is to transform a given coherent state $\rho$ into a set of $n$-level maximally coherent states via SIO. By optimal we mean that the average coherence of the final ensemble is maximal. This protocol, inspired by an entanglement distillation protocol in Refs. \cite{Jonathan,Hardy}, was first proposed in Ref. \cite{Torun}. In that paper, with the help of the $l_1$ norm of coherence, Torun \emph{et al.} studied this coherence distillation protocol for pure states. However, since we often encounter mixed states rather than pure states, an immediate question is how to extend this protocol to the mixed state case.

In this paper, we solve this problem completely by developing the optimal probabilistic coherence distillation protocol in the mixed state case. To obtain this result, we present the necessary and sufficient condition for the transformation from a mixed state into a pure-state ensemble via SIO. With this condition and the simplex algorithm, we then accomplish the optimal probabilistic coherence distillation protocol by presenting an analytical expression of the maximal average distillable coherence for any state and constructing the corresponding operation achieving this bound. Our protocol is universal since it can be applied to any coherence measure.

\section{Resource theory of coherence}
To present our result clearly, we recall some elementary notions of the resource theory of coherence. Let $\{\ket{i}\}_{i=1}^d$ be the prefixed basis in the finite dimensional Hilbert space. A state is said to be incoherent if it is diagonal in the basis and the set of such states is denoted by $\mathcal{I}$. Coherent states are those not of this form. For a pure state $\ket{\varphi}$, we will write $\varphi:=\ket{\varphi}\bra{\varphi}$. The $d$-dimensional maximally coherent state has the form $\ket{\psi^d}=\frac1{\sqrt{d}}\sum_{i=1}^{d}\ket{i}$ \cite{Baumgratz} and we will denote
\begin{eqnarray}
\ket{\psi^n}=\frac1{\sqrt{n}}\sum_{i=1}^{n}\ket{i}, ~\text{for}~ 1\leq n\leq d
\end{eqnarray}
as an $n$-level maximally coherent state. The SIO is a completely positive trace preserving (CPTP) map expressed as $
\Lambda(\rho)=\sum_{\mu=1}^N K_\mu\rho K_\mu^\dagger$,
where the Kraus operators $K_\mu$ satisfy not only $\sum_{\mu=1}^N K_\mu^\dagger K_\mu= I$ but also $K_\mu\mathcal{I}K_\mu^\dagger\subset \mathcal{I}~\text{and}~K_\mu^\dag\mathcal{I}K_\mu\subset \mathcal{I}$ for every $K_\mu$ \cite{Winter, Yadin}. One sees by inspection that there is at most one nonzero element in each column and row of $K_\mu$, and $K_\mu$ are called SIO Kraus operators. From this, it is elementary to show that a projector is incoherent if it has the form $\mathbb{P}_\mu=\sum_{i}\ket{i}\bra{i}$ and we will denote with $\mathbb{P}_\mu$ a generic strictly incoherent projector. A CPTP map is an incoherent operation (IO) if each $K_\mu$ only satisfies $K_\mu^\dag\mathcal{I}K_\mu\subset \mathcal{I}$ for all $\mu$ \cite{Baumgratz}. A stochastic SIO \cite{Liu1}, denoted as $\Lambda_s(\rho)$, is defined as $$
\Lambda_s(\rho)=\frac{\sum_{\mu=1}^L K_\mu\rho K_\mu^{\dagger}}{\Tr(\sum_{\mu=1}^LK_\mu\rho K_\mu^{\dagger})},$$
where $\{K_{1},K_{2},\dots, K_{L}\}$ satisfies $\sum_{\mu=1}^L K_\mu^{\dagger}K_{\mu}\leq I$. Clearly, the state $\Lambda_s(\rho)$ is obtained with probability $P=\Tr(\sum_{\mu=1}^LK_\mu\rho K_\mu^{\dagger})$. SIO correspond to the case $P=1$. We will use $\Delta(\rho)=\sum_i\ket{i}\bra{i}\rho\ket{i}\bra{i}$ to denote the fully dephasing channel.

A functional $C$ can be taken as a coherence measure if it satisfies the four conditions \cite{Baumgratz,Yadin}: (C1) the coherence being zero for incoherent states; (C2) the monotonicity of coherence under SIO; (C3) the monotonicity of coherence under selective measurements on average; and (C4) the non-increasing of coherence under mixing of quantum states. Based on these conditions, various coherence measures have been put forward. Let us recall the relative entropy of coherence $C_r$ \cite{Baumgratz}, the $l_1$ norm of coherence $C_{l_1}$ \cite{Baumgratz}, and the coherence rank $C_R$ \cite{Winter}, which will be used in this paper. Here, $C_r(\rho)=S(\Delta\rho)-S(\rho)$, where $S(\rho)=-\Tr(\rho\ln\rho)$ is the von Neumann entropy (the base of our logarithms being 2). The $l_1$ norm of coherence is defined as $C_{l_1}(\rho)=\sum_{i\neq j}\abs{\rho_{ij}}$. The coherence rank $C_R$ of a pure state (not necessarily normalized), $\ket{\varphi}=\sum_{i=1}^Rc_i\ket{i}$ with $c_i\neq 0$, is defined as the number of nonzero terms in this decomposition minus $1$ \cite{note}, i.e., $C_R(\varphi)=R-1$.

\section{Transformation criterion}
The optimal probabilistic coherence distillation is concerned with the transformation from a general state into a special pure-state ensemble via SIO. We thus start by giving the following theorem which provides the necessary and sufficient conditions for such transformations.

Theorem 1. The transformation $\rho\to\{p_{\mu n},\ket{\varphi_{\mu n}}\}_n^\mu$ can be realized via SIO if and only if  there is an orthogonal and complete set of incoherent projectors $\{\mathbb{P}_\mu\}$ such that, for all $\mu$, there are
\begin{eqnarray}\label{condition}
\frac{\mathbb{P}_\mu\rho\mathbb{P}_\mu}{\Tr(\mathbb{P}_\mu\rho\mathbb{P}_\mu)}=\psi_\mu~\text{and}~\Delta\psi_\mu\prec\sum_np_{n|\mu}\Delta\varphi_{\mu n}^\downarrow,
\end{eqnarray}
where $\psi_\mu$ are pure states, $p_\mu=\Tr(\mathbb{P}_\mu\rho\mathbb{P}_\mu)$,  and $p_{n|\mu}:=p_{\mu n}/p_\mu$.

Here, let $\rho$ and $\sigma$ be two states with their eigenvalues being $(\uplambda_1\geq\uplambda_2\geq\cdots\geq\uplambda_d)$ and $(\uplambda_1^\prime\geq\uplambda_2^\prime\geq\cdots\geq\uplambda_d^\prime)$, respectively. Hereafter, we will take the eigenvalues of a state in this descending order. Then, the majorization relation $\rho\prec\sigma$ implies that there are $C_k(\rho):=\sum_{i=k}^d\uplambda_i\geq C_k(\sigma):=\sum_{i=k}^d\uplambda_i^\prime$ for all $1\leq k\leq d-1$ \cite{Bhatia} and we denote $\rho\prec\sum_np_n\rho_n^\downarrow$ as $C_k(\rho)\geq\sum_np_nC_k(\rho_n)$ for all $1\leq k\leq d-1$.

\emph{Proof.} We first prove the \emph{if} part of the theorem, i.e., if  $\rho$ satisfies the conditions in Eq. (\ref{condition}), then we can transform $\rho$ into the ensemble $\{p_{\mu n},\ket{\varphi_{\mu n}}\}$ via SIO.

To this end, we act $\Lambda_1(\cdot)=\sum_\mu\mathbb{P}_\mu(\cdot)\mathbb{P}_\mu$ on $\rho$ with $\{\mathbb{P}_\mu\}$ being an orthogonal and complete set of incoherent projectors. After that, we obtain \begin{eqnarray}\nonumber
\Lambda_1(\rho)=\sum_\mu\mathbb{P}_\mu\rho\mathbb{P}_\mu=\bigoplus_\mu p_\mu\psi_\mu,
\end{eqnarray}
where $p_\mu=\Tr(\mathbb{P}_\mu\rho\mathbb{P}_\mu )$. Then, by using the condition $\Delta\psi_\mu\prec\sum_np_{n|\mu}\Delta\varphi_{\mu n}^\downarrow$ and the result in Ref. \cite{Chitambar3}, which says that the transformation from a pure state $\psi$ into an ensemble $\{p_n,\varphi_n\}$ can be realized via  $\{K_n\}_n$ which forms a SIO if and only if there are $\Delta\psi\prec\sum_np_n\Delta\varphi_n^\downarrow$, it is direct to obtain that we can always find some SIO $\{K_n^\mu\}_n$ realizing the transformation from $\psi_\mu$ into $\{p_{n|\mu},\ket{\varphi_{\mu n}}\}$. Therefore, we can achieve the transformation from $\rho$ into $\{p_{\mu n},\ket{\varphi_{\mu n}}\}$ by first performing $\{\mathbb{P}_\mu\}_\mu$ on $\rho$ and then, according to the index $\mu$, performing $\{K_n^\mu\}_n$ on it. This completes the $\emph{if}$ part of the theorem.

Next, we prove the \emph{only if} part of the theorem, i.e., if $\rho$ can be transformed into the ensemble $\{p_{\mu n},\ket{\varphi_{\mu n}}\}_n^\mu$ via some SIO $\{K_n^\mu\}_n^\mu$, then $\rho$ should satisfy the conditions in Eq. (\ref{condition}).

To this end, suppose the transformation from the state $\rho$ into the ensemble $\{p_{\mu n},\ket{\varphi_{\mu n}}\}_n^\mu$ can be realized via $\{K_n^\mu\}_n^\mu$. It implies that there is a stochastic SIO $\Lambda_s^{\{\mu n\}}(\cdot)$ which is constructed by a subset of  $\{K_n^\mu\}$ such that \begin{eqnarray}
\Lambda_s^{\{\mu n\}}(\rho)=p_{\mu n}\varphi_{\mu n}\end{eqnarray}
for all $\mu,n$, where $p_{\mu n}=\Tr\left(\Lambda_s^{\{\mu n\}}(\rho)\right)$. From the result in Ref. \cite{Liu3}, which says that a pure coherent state $\ket{\varphi}$ can be obtained from a mixed state $\rho$ via stochastic SIO if and only if there is an incoherent projector $\mathbb{P}$ with  the coherence rank of $\mathbb{P}\rho\mathbb{P}$ being greater than or equal to that of $\ket{\varphi}$, we immediately obtain that it is the parts of $\rho$ such that $\mathbb{P}\rho\mathbb{P}$ being rank one are useful in this transformation. This implies that each $\varphi_{\mu n}$ was obtained from these parts of $\rho$, i.e., the part such that the rank of $\mathbb{P}\rho\mathbb{P}$ is one. Then, it is direct to see that a pure-state ensemble $\{p_{\mu n},\ket{\varphi_{\mu n}}\}_n^\mu$ can be obtained from $\rho$ via a SIO $\{K_n^\mu\}$  if and only if the same ensemble can be obtained from some state
\begin{eqnarray}
\widetilde{\rho}:=\sum_\mu\mathbb{P}_\mu\rho\mathbb{P}_\mu=\bigoplus_\mu p_\mu\psi_\mu
\end{eqnarray}
 via the same $\{K_n^\mu\}$, where $\{\mathbb{P}_\mu\}$  is an orthogonal and complete set of incoherent projectors, $p_\mu=\Tr(\mathbb{P}_\mu\rho\mathbb{P}_\mu)$, and each $\mathbb{P}_\mu\rho\mathbb{P}_\mu=p_\mu\psi_\mu$ is rank one.

Thus to realize the transformation from the state $\rho$ into the ensemble $\{p_{\mu n},\ket{\varphi_{\mu n}}\}_n^\mu$ via $\{K_n^\mu\}$, we only need to consider the transformation $\widetilde{\rho}\to\{p_{\mu n},\ket{\varphi_{\mu n}}\}_n^\mu$. The transformation from $\widetilde{\rho}$ into $\{p_{\mu n},\ket{\varphi_{\mu n}}\}_n^\mu$ via $\{K_n^\mu\}$ further implies that there are \begin{eqnarray}
K_n^\mu\ket{\psi_\mu}\bra{\psi_\mu}{K_n^\mu}^\dag=p_{n|\mu}^\prime\varphi_{\mu n},
\end{eqnarray}
where $p_{n|\mu}^\prime=\Tr(K_n^\mu\ket{\psi_\mu}\bra{\psi_\mu}{K_n^\mu}^\dag)$. Let the ensemble obtained after performing $\{K_n^\mu\}$ on $\ket{\psi_\mu}$ be $\{p_{n|\mu},\ket{\varphi_{\mu n}}\}_n$. Noting that the transformation from a pure state $\psi$ into the ensemble $\{p_n,\varphi_n\}$ can be realized via a SIO $\{K_n\}$ if and only if there are $\Delta\psi\prec\sum_np_n\Delta\varphi_n^\downarrow$, we immediately obtain that, to realize the transformation from $\rho$ into the ensemble $\{p_{\mu n},\ket{\varphi_{\mu n}}\}_n^\mu$ with $p_{\mu n}=p_\mu p_{n|\mu}$ via $\{K_n^\mu\}$, $\rho$ should satisfy the conditions in Eq. (\ref{condition}). This completes the \emph{only if} part of the theorem.

\section{Probabilistic coherence distillation}
With Theorem 1, let us move to consider the protocol of the optimal probabilistic coherence distillation, which can be formally defined as follows:~~
Given a state $\rho$, the aim of the protocol is to transform it into a set of $n$-level maximally coherent states via SIO (as illustrated in Fig.1), that is, we want to accomplish the transformation
\begin{eqnarray}
\rho\stackrel{\text{SIO}}{\longrightarrow}\{p_n,\ket{\psi^n}\}
\end{eqnarray}
with the coherence of the final ensemble $\{p_n,\ket{\psi^n}\}$, $\overline{C}=\sum_np_nC(\psi^n)$, as large as possible. Here, $C$ is an arbitrary coherence measure fulfilling (C1)-(C4).
\begin{figure}[ht]
 \centering
 \includegraphics[height=3.5cm]{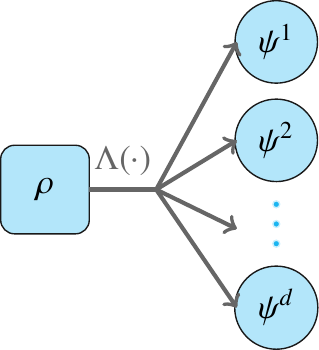}
 \caption{Optimal probabilistic distillation of quantum coherence. This figure illustrates the protocol that transforms a general state $\rho$ into a set of  $n$-level maximally coherent states $\ket{\psi^n}$ with $n=1,\cdots,d$ via SIO $\Lambda(\cdot)$.}
\end{figure}

Based on the result above, we present the protocol of obtaining $\overline{C}_{\max}(\rho)$ as the following three steps:

First, identify the pure coherent-state subspaces of $\rho$.

From Theorem 1, we note that it is the parts of $\rho$ such that $\mathbb{P}\rho\mathbb{P}$ being rank one are useful in the transformation $\rho\to\{p_{\mu},\ket{\varphi_{\mu}}\}_\mu$. For the sake of simplicity, we call these parts as the pure coherent-state subspaces of $\rho$. More precisely, if there is an incoherent projector $\mathbb{P}$ such that $\mathbb{P}\rho\mathbb{P}= \varphi$ with the coherence rank of $\varphi$ being $n\geq0$, then we say that $\rho$ has an $n+1$-dimensional pure coherent-state subspace corresponding to $\mathbb{P}$. Furthermore, we say that the pure coherent-state subspaces with the projector $\mathbb{P}$ for $\rho$ are maximal if the pure coherent-state subspace cannot be expanded to a larger one with a incoherent projector $\mathbb{P}^{\prime}$ such that $\mathbb{P}^{\prime}\rho\mathbb{P}^{\prime}= \varphi^{\prime}$, $\varphi^{\prime}\neq\varphi$, and $\mathbb{P}\varphi^{\prime}\mathbb{P}= \varphi$.  The pure coherent-state subspaces of $\rho$ can be identified by using
\begin{eqnarray}
\mathcal{A}=(\Delta\rho)^{-\frac12}\abs{\rho}(\Delta\rho)^{-\frac12}\label{A},
\end{eqnarray}
which was presented in Ref. \cite{Liu1}.
Here, for $\rho=\sum_{ij}\rho_{ij}\ket{i}\bra{j}$, $\abs{\rho}$ reads $|\rho|=\sum_{ij}|\rho_{ij}|\ket{i}\bra{j}$ and
$(\Delta\rho)^{-\frac12}$ is the diagonal matrix with elements
$(\Delta\rho)^{-\frac12}_{ii}= \left\{
  \begin{array}{ll}
     \rho_{ii}^{-\frac12}, &\text{if} ~ \rho_{ii}\neq0;\\
    0,&\text{if}~ \rho_{ii}= 0.
  \end{array}\right.$
It can be shown that $\mathbb{P}\rho\mathbb{P}$ is rank $1$ if and only if all its corresponding elements of $\mathcal{A}$ are $1$ \cite{Liu1}. From this, we can obtain that if there are $n$-dimensional principal submatrices $\mathcal{A}_\mu$ of $\mathcal{A}$ with all its elements being $1$, then the corresponding subspace of $\rho$ is an $n$-dimensional pure coherent-state subspace. By using this result, one can easily identify the pure coherent-state subspaces of $\rho$. For a state $\rho$, let the corresponding Hilbert subspaces of principal submatrices $\mathcal{A}_\mu$ ($\mu=1,\cdots,\mathcal{U}$) be $\mathcal{H}_\mu$, which is spanned by
$\{\ket{i_1^\mu},\ket{i_2^\mu},\cdots,\ket{i_{d_\mu}^\mu}\}\subset\{\ket{1}, \ket{2},\cdots,\ket{d}\}$ and the corresponding incoherent projectors be $\mathbb{P}_\mu$, with its rank being $d_\mu$, i.e., $
  \mathbb{P}_\mu=\ket{i^\mu_1}\bra{i^\mu_1}+\ket{i^\mu_2}\bra{i^\mu_2}+\cdots+\ket{i^\mu_{d_\mu}}\bra{i^\mu_{d_\mu}}$.
Acting $\{\mathbb{P}_\mu\}$ on the state $\rho$, we then obtain the set $\{\varphi_\mu\}_{\mu=1}^\mathcal{U}$, where $\varphi_\mu$ have the form
$\varphi_\mu=(\mathbb{P}_\mu\rho\mathbb{P}_\mu)/\Tr(\mathbb{P}_\mu\rho\mathbb{P}_ \mu)$.
Let the pure states corresponding to maximal pure coherent-state subspaces be
\begin{eqnarray}
\frac{\mathbb{P}^\text{m}_\mu\rho\mathbb{P}^\text{m}_\mu}{\Tr(\mathbb{P}^\text{m}_\mu\rho\mathbb{P}^\text{m}_\mu)}=\varphi^\text{m}_\mu.
\end{eqnarray}
Here, $\mathbb{P}^\text{m}_\mu$ are the incoherent projectors corresponding  to maximal pure coherent-state subspaces.
Then, after acting the incoherent projectors $\{\mathbb{P}^\text{m}_\mu\}$ on $\rho$, we obtain a set of pure states $\varphi^\text{m}_\mu$ with probability $p_\mu=\Tr(\mathbb{P}^\text{m}_\mu\rho\mathbb{P}^\text{m}_\mu)$, i.e., there is
\begin{eqnarray}
\Lambda_\mathbb{P}(\rho)=\sum_{\mu=1}^\mathcal{U}\mathbb{P}^\text{m}_\mu\rho\mathbb{P}^\text{m}_\mu=\bigoplus_{\mu=1}^\mathcal{U} p_\mu\varphi_\mu^\text{m}. \label{mixed_first}
\end{eqnarray}
By Theorem 1 and  the definition of $\{\mathbb{P}^\text{m}_\mu\}$ and $\{\mathbb{P}_\mu\}$, it is straightforward to see that, to obtain the optimal probabilistic distillation protocol, we only need to consider the state $\rho^\text{m}=\bigoplus_{\mu=1}^\mathcal{U} p_\mu\varphi_\mu^\text{m}$,  since general $\rho^\prime=\bigoplus_{\mu}p_\mu\varphi_\mu$ can be obtained from $\rho^m$ via SIO.

Second, calculate $\overline{C}_{\max}(\varphi_\mu^\text{m})$ using the simplex algorithm.

The results presented in the first step imply that, to obtain $\overline{C}_{\max}(\rho)$, we only need to calculate each $\overline{C}_{\max}(\varphi_\mu^\text{m})$ since there is $\overline{C}_{\max}(\rho)=\sum_{\mu=1}^\mathcal{U} p_\mu\overline{C}_{\max}(\varphi_\mu^\text{m})$. To this end, we note that there is a constraint in the transformation $\ket{\varphi_\mu^\text{m}}\to\{p_n,\ket{\psi^n}\}$, i.e., $\ket{\varphi_\mu^\text{m}}\to\{p_n,\ket{\psi^n}\}$ can be realized via SIO if and only if there are $\Delta\psi\prec\sum_np_n\Delta\varphi_n^\downarrow$. These conditions can be written as $\sum_{i=l}^d\abs{\varphi_i}^2\geq\sum_{n=l}^dp_n\left(\frac{n-l+1}{n}\right)$, where $l\leq n$. This implies that the problem of finding $\overline{C}_{\max}(\varphi_\mu^\text{m})$ can be reformulated as the following linear programming problem
\begin{eqnarray}
\text{maximum}~~&&\bf{c}^T\bf{p}\nonumber\\
\text{subject~to}~~  &&\text{A}\bf{p}\leq \bf{q}, ~~\textbf{p}\geq\textbf{0}.   \label{linear}
\end{eqnarray}
Here, with the superscript $T$ being transpose operation, the vectors $\textbf{p}$ and $\textbf{q}$ read $\textbf{p}=(p_1,p_2,\dots,p_d)^T$ and $\textbf{q}=\left(C_1(\Delta\varphi), C_2(\Delta\varphi), \cdots, C_d(\Delta\varphi)\right)^T$, respectively. The matrix $A$ reads $A=\sum_{l\leq n}\frac{n+1-l}{n}\ket{n}\bra{l}$ and the $\textbf{c}=\left(f(1),f(2),\cdots,f(d)\right)^T$, where $f(n)$ is the value of $\ket{\psi^n}$ under some specific coherence measure and it is direct to obtain $f(n)\geq f(n-1)\geq\cdots\geq f(2)> f(1)=0$.

The optimization problem presented in Eq. (\ref{linear}) can be efficiently solved by using the simplex algorithm \cite{Luenberger}, that is, we can always obtain the maximum $\overline{C}_{\max}(\varphi_\mu^\text{m})=\text{max}_{\bf{p}}~\bf{c}^T\bf{p}$. Hence, we can present the analytical expression $\overline{C}_{\max}(\rho)=\sum_{\mu=1}^\mathcal{U} p_\mu\overline{C}_{\max}(\varphi_\mu^\text{m})$ of $\rho$. Since the corresponding operation achieving this bound can be constructed  by Theorem 1, the protocol presented here is universal. Here we consider the case that $nf(n)$ is convex, which is satisfied by various coherence measures such as $C_r$ and $C_{l_1}$. The generalization to others is straightforward. For these coherence measures, we obtain the optimal solution as (see Appendix for further technical details)
\begin{eqnarray}
\overline{C}_{\max}(\varphi_\mu^\text{m})=\sum_{i=1}^di(\abs{c_i}^2-\abs{c_{i+1}}^2)f(i). \label{value}
\end{eqnarray}

Third, present $\overline{C}_{\max}(\rho)$ and the corresponding operation.

Let $\ket{\varphi^\text{m}_\mu}=\sum_{i=1}^{d_\mu}c^\mu_i\ket{i}$. The transformation  $\ket{\varphi_\mu^\text{m}}\to\{p_n,\ket{\psi^n}\}$ can be realized via the SIO
$\Lambda_\mu(\cdot)= \sum_{n=1}^{d_\mu}K_n^\mu(\cdot)K_n^\mu$,
where $K_n^\mu=\sqrt{\frac{p_n}{n}}\left(\sum_{i=1}^{n}\frac{\ket{i}\bra{i}}{c^\mu_i}\right)$
with $p_{d_\mu}=d_\mu\abs{c_{d_\mu}^\mu}^2$ and $p_n=n(\abs{c_n^\mu}^2-\abs{c_{n+1}^\mu}^2)$ for $n=1,\cdots,d_\mu-1$.
With the aid of $\Lambda_\mu(\cdot)$, we then obtain the operation corresponding to the optimal distillation with the form $
\Lambda(\cdot)=\bigoplus_{\mu=1}^\mathcal{U}\Lambda_\mu(\cdot)$.
As it is straightforward to examine that $\sum_{n=1}^{d_\mu}{K_n^\mu}^\dag K_n^\mu=I_{d_\mu}$ and each $K_n^\mu$ has at most one nonzero element in each column and row,  $\Lambda(\cdot)$ is a SIO. Since we can transform $\rho$ into $\psi^n$ with probability $p_n=\sum_{\mu=1}^{\mathcal{U}} p_\mu\left(n(\abs{c_n^\mu}^2-\abs{c_{n+1}^\mu}^2)\right)$ in the above protocol, the maximal average distilled coherence of $\rho$ is $
\overline{C}_{\max}(\rho)=\sum_{\mu=1}^\mathcal{U}\sum_{n=1}^{d_\mu} p_\mu\left(\abs{c_n^\mu}^2-\abs{c_{n+1}^\mu}^2\right)nf(n)$.

With the coherence measures considered here, i.e., $nf(n)$ is convex, we then summarize the above results as follows.

Theorem 2. For state $\rho$, let the state corresponding to its maximal pure coherent-state subspaces be $\Lambda_\mathbb{P}(\rho)=\bigoplus_{\mu=1}^\mathcal{U} p_\mu\varphi_\mu^{\text{m}}$. Then, the maximal average distillable coherence of $\rho$ is
\begin{eqnarray}
\overline{C}_{\max}(\rho)=\sum_{\mu=1}^\mathcal{U}\sum_{n=1}^{d_\mu} p_\mu\left(\abs{c_n^\mu}^2-\abs{c_{n+1}^\mu}^2\right)nf(n).
\end{eqnarray}
The corresponding SIO is \begin{eqnarray}
\Lambda(\cdot)=\bigoplus_{\mu=1}^\mathcal{U}\left(\sum_{n=1}^{d_\mu}K_n^\mu(\cdot)K_n^\mu\right),
\end{eqnarray}
where $K_n^\mu=\sqrt{\frac{p_n}{n}}\left(\sum_{i=1}^{n}\frac{\ket{i}\bra{i}}{c^\mu_i}\right)$
with $p_{d_\mu}=d_\mu\abs{c^\mu_{d_\mu}}^2$ and $p_n=n(\abs{c_n^\mu}^2-\abs{c_{n+1}^\mu}^2)$ for $n=1,\cdots,d_\mu-1$.

In particular, if the initial state is a pure one and the coherence measure is the $C_{l_1}(\rho)$, i.e., $f(n)=C_{l_1}(\psi^n)=n-1$, then we can obtain the results in Ref. \cite{Torun}.

\section{Further Discussions}

Before concluding, we give the following remarks concerning the above distillation protocol.

The first is the irreversibility of the distillation protocol above. Here, irreversibility means that we cannot convert a state into another and then recover the original state with certainty. We can see the irreversibility of this protocol from two aspects. On the one hand, when we consider the pure state case, the irreversible was presented by using $C_{l_1}$  \cite{Torun}. On the other hand, we obtain that, for a mixed state $\rho$, there is another source of  irreversibility result from $\chi:=\left(\rho-\bigoplus_\mu p_\mu\varphi_\mu^{\text{m}}\right)$.  From Theorem 1, we see that this part is completely consumed in the above distillation protocol. It is not hard to show that a state $\rho$ is reversible in this protocol if and only if $\rho=\bigoplus_\mu p_\mu\varphi_\mu$ with each $\varphi_\mu$ being an $n$-level maximally coherent state.

The second is the relation between distillable states in the protocol here and the distillable states in asymptotic regime via SIO. We say a state  $\rho$ is distillable in the protocol here if any $n(\geq2)$-level maximally coherent state can be obtained from it with nonzero probability. A state is distillable in asymptotic regime if we can distill pure coherence from it \cite{Lami,Lami1,Zhao}. It is proved that a state is distillable in the asymptotic regime if it contains any rank-one submatrix with its coherence rank greater than or equal to 1. By Theorems 1 and 2, it is direct to obtain that the set of distillable states in the protocol here and the set of distillable states in asymptotic regime are identical.

The third one is that there is an operational gap between IO and SIO in this distillation protocol. To show this, let us consider the state \cite{Liu4}\\~\\
$
  ~~~~~~~~~~~~~~~~~~~~~~~~\rho=\begin{pmatrix}
    \frac14&0&\frac1{2\sqrt{5}}&\frac1{4\sqrt{5}}\\
    0&\frac14&-\frac1{4\sqrt{5}}&\frac1{2\sqrt{5}}\\
   \frac1{2\sqrt{5}}&-\frac1{4\sqrt{5}}&\frac14&0\\
    \frac1{4\sqrt{5}}&\frac1{2\sqrt{5}}& 0&\frac14
  \end{pmatrix}.
$\\~\\
It is straightforward to examine that there is not an incoherent projector $\mathbb{P}$ such that the coherence rank of $\mathbb{P}\rho\mathbb{P}$ is greater than or equal to 1. From Theorem 1, we cannot obtain any $\psi^n$ ($n\geq2$) with nonzero probability from it via SIO. However, this is not the case when we consider IO. To see this, let the operation be $\Lambda(\cdot)=K_1(\cdot)K_1^\dag+K_2(\cdot)K_2^\dag$, where~\\~\\
$~~~~~~~~~~~~~~K_1=\begin{pmatrix}
    \frac45&\frac35&0&0\\
    0&0&\frac1{\sqrt{5}}&\frac2{\sqrt{5}}\\
    0& 0&0&0\\
    0& 0& 0&0
  \end{pmatrix},~~
 K_2=\begin{pmatrix}
    -\frac35&\frac45&0&0\\
    0&0&-\frac2{\sqrt{5}}&\frac1{\sqrt{5}}\\
    0& 0&0&0\\
    0& 0& 0&0
  \end{pmatrix}.$\\~\\
By acting $\Lambda(\cdot)$ on $\rho$, we obtain that $\Lambda(\rho)=\psi^2$. This means that $\overline{C}(\rho)_{\max}\geq f(2)>0$. This example implies that IO are stronger than SIO when we consider the above distillation protocol.

\section{Conclusions}
In summary, we fully characterized the optimal probabilistic coherence distillation protocol, whose aim is to transform a coherent state into a set of $n$-level maximally coherent states via SIO. To this end, we have presented the necessary and sufficient condition for the transformation from a mixed state into a pure-state ensemble via SIO. With the aid of this condition and the simplex algorithm, we could accomplish the optimal probabilistic coherence distillation protocol by presenting an analytical expression of the maximal average distillable coherence for a general state and constructing the corresponding operation achieving this bound. Interestingly, our protocol can be applied to any coherence measure. Our protocol provides a practical protocol for efficient quantum coherence manipulation of mixed states.

\section*{Acknowledgments}
This work is supported by the National Natural Science Foundation of China (NSFC) (Grants No. 12088101, No. U1930403, and No. U1930402). C.L.L acknowledges support from the China Postdoctoral Science Foundation Grant No. 2021M690324.

\section*{Appendix}

Here we give the details of obtaining $\overline{C}_{\max}(\varphi_\mu^\text{m})$ in Eq. (\ref{value}).

To this end, we first convert the optimization problem in Eq. (\ref{value}) into its standard form. This means that we transform all the constraint conditions in Eq. (\ref{value}) into equalities by introducing slack variables $p_s\geq0$ with $d+1\leq s\leq 2d$, i.e., transform $a_{i1}p_1+a_{i2}p_2+\cdots+a_{id}p_d\leq C_k(\Delta\varphi)$ into $a_{i1}p_1+a_{i2}p_2+\cdots+a_{id}p_d+p_{s}= C_k(\Delta\varphi)$ with the aid of $p_s$, where $p_s\geq0$. The problem in Eq. (\ref{value}) can be then reexpressed as
\begin{eqnarray}
\text{maximum}~~&&{\textbf{c}^\prime}^T\textbf{p}^\prime\nonumber\\
\text{subject~to}~~  &&\text{B}\textbf{p}^\prime= \textbf{q},~~\textbf{p}^\prime\geq\textbf{0}\label{linear1}
\end{eqnarray}
where the vector $\textbf{c}^\prime=\left(f(1),f(2),\cdots,f(d),0,0,\cdots,0\right)^T$, the matrix
\begin{eqnarray}
\text{B}&&=\begin{pmatrix}
    1&1&1&\cdots&1&1&0&0&\cdots&0\\
    0&\frac12&\frac23&\cdots&\frac{d-1}d&0&1&0&\cdots&0\\
    0&0&\frac13&\cdots&\frac{d-2}d&0&0&1&\cdots&0\\
    \vdots&\vdots&\vdots&\ddots&\vdots&\vdots&\vdots&\vdots&\ddots&\vdots\\
    0&0&0&\cdots&\frac1d&0&0&0&\cdots&1
  \end{pmatrix},\label{matrix1}
\end{eqnarray} and $\textbf{p}^\prime=(p_1,p_2,\dots,p_d,p_{d+1},\cdots,p_{2d})^T$. The standard form in  Eq. (\ref{value}) can be further converted to the Table I. In Table I, $\textbf{b}_j$ is the $j$-th column of the matrix $\text{B}$ in Eq. (\ref{value}), the leftmost column indicates the original variables, the next $2d$ columns contain the coefficients for the $d$ original variables and $d$ slack variables, and the rightmost column is the constants $q_i$.
\begin{table}[h!]
  \begin{center}
    \caption{Initial table of the linear programming problem in Eq. (\ref{value}) with   $p_{d+1},\cdots,p_{2d}$ being basic variables and $p_{1},\cdots,p_{d}$ being free variables.}
    \begin{tabular}{c|c|c|c|c|c|c|c|c|c|c|c}
    \hline
     &$ \textbf{b}_1$ & $ \textbf{b}_2$ & $ \textbf{b}_3$& $\cdots$ & $ \textbf{b}_d$& $ \textbf{b}_{d+1}$ &$ \textbf{b}_{d+2}$&$\textbf{b}_{d+3}$&$\cdots$&
     $ \textbf{b}_{2d}$&$ \textbf{q}$\\
      \hline
    $p_{d+1}$& 1&1&1&$\cdots$&1&1&0&0&$\cdots$&0&$q_1$\\  \hline
     $p_{d+2}$&0&$\frac12$&$\frac23$&$\cdots$&$\frac{d-1}d$&0&1&0&$\cdots$&0&$q_2$\\  \hline
     $p_{d+3}$&0&0&$\frac13$&$\cdots$&$\frac{d-2}d$&0&0&1&$\cdots$&0&$q_3$\\  \hline
    $\vdots$&$\vdots$&$\vdots$&$\vdots$&$\ddots$&$\vdots$&$\vdots$&$\vdots$&$\vdots$&$\ddots$&$\vdots$&$\vdots$\\  \hline
    $p_{2d}$& 0&0&0&$\cdots$&$\frac1d$&0&0&0&$\cdots$&1&$q_d$\\
    \hline
   c& $-f(1)$&$-f(2)$&$-f(3)$&$\cdots$&$-f(d)$&0&0&0&$\cdots$&0&0\\
    \hline
    \end{tabular}
  \end{center}
\end{table}
We first take the variables $p_{d+1},\cdots, p_{2d}$ as basic variables, which presented in the leftmost column and the remaining variables $ p_{1},\cdots,p_{d}$ as free variables. A feasible solution of Eq. (\ref{linear1}) is a vector $\textbf{p}^\prime$ fulfilling both $\text{B}\textbf{p}^\prime= \textbf{q}$ and $\textbf{p}^\prime\geq\textbf{0}$.

As the first step of the simplex algorithm, we set the basic variables to $p_{d+i}=q_i$ for $i=1,\cdots, d$ and set all the free variables to $p_i=0$. It is obvious that the value of the objective function is $0$ at this time. Next, by carrying out Gaussian elimination, we can generate a new feasible solution from the old one by replacing a basic variable with a free variable while preserving the nonnegativity of the solution \cite{note2}. For example, if there is $q_1\geq2q_2$, we can then transform $p_2$ from a free variable into a basic variable and $p_{d+2}$ into a free variable. Then, by direct calculation, we obtain Table II. In Table II, there are $c_i=-f(i)+(\frac{i-1}i)f(2)$ with $i=3,\cdots,d$.
\begin{table}[h!]
  \begin{center}
    \caption{The table with $p_{d+1},p_2,p_{d+3},\cdots,p_{2d}$ being the basic variables and $p_1,p_{d+2},p_3,\cdots,p_d$ being free variables. }
    \begin{tabular}{c|c|c|c|c|c|c|c|c|c|c|c}
    \hline
     &$ \textbf{b}_1$ & $ \textbf{b}_2$ & $ \textbf{b}_3$& $\cdots$ & $ \textbf{b}_d$& $ \textbf{b}_{d+1}$ &$ \textbf{b}_{d+2}$&$\textbf{b}_{d+3}$&$\cdots$&
     $ \textbf{b}_{2d}$&$ \textbf{q}$\\
      \hline
    $p_{d+1}$& 1&0&$-\frac13$&$\cdots$&$\frac{2-d}{d}$&1&-2&0&$\cdots$&0&$q_1-2q_2$\\  \hline
     $p_{2}$&0&$1$&$\frac43$&$\cdots$&$\frac{2(d-1)}d$&0&2&0&$\cdots$&0&$2q_2$\\  \hline
     $p_{d+3}$&0&0&$\frac13$&$\cdots$&$\frac{d-2}d$&0&0&1&$\cdots$&0&$q_3$\\  \hline
    $\vdots$&$\vdots$&$\vdots$&$\vdots$&$\ddots$&$\vdots$&$\vdots$&$\vdots$&$\vdots$&$\ddots$&$\vdots$&$\vdots$\\  \hline
    $p_{2d}$& 0&0&0&$\cdots$&$\frac1d$&0&0&0&$\cdots$&1&$q_d$\\
    \hline
   c& $0$&$0$&$c_3$&$\cdots$&$c_d$&0&2&0&$\cdots$&0&$2q_2$\\
    \hline
    \end{tabular}
  \end{center}
\end{table}
We note that the basic variables become $p_{d+1},p_2,p_{d+3},\cdots,p_{2d}$. The value of the objective function then becomes $2q_2$. The procedure continues until all elements $c_i$ with $i=1,\cdots, 2d-1$ of the last lines being nonnegative, which is a sufficient condition for obtaining the maximum value in  Eq. (\ref{value}) \cite{Luenberger}. Suppose $nf(n)$ is convex, which is satisfied by various coherence measures such as $C_r$ and $C_{l_1}$. The generalization to others is straightforward by using the similar method. For the problem in Eq. (\ref{linear1}), let us consider a special case where $p_1,p_2,\cdots,p_d$ are turned into basic variables. By direct calculation, we then  obtain Table III.
\begin{table}[h!]
  \begin{center}
    \caption{The final table with $p_{1},\cdots,p_{d}$ being basic variables and $p_{d+1},\cdots,p_{2d}$ being free variables.}
    \begin{tabular}{c|c|c|c|c|c|c|c|c|c|c|c}
    \hline
     &$ \textbf{b}_1$ & $ \textbf{b}_2$ & $ \textbf{b}_3$& $\cdots$ & $ \textbf{b}_d$& $ \textbf{b}_{d+1}$ &$ \textbf{b}_{d+2}$&$\textbf{b}_{d+3}$&$\cdots$&
     $ \textbf{b}_{2d}$&$ \textbf{q}$ \\
      \hline
    $p_1$& 1&0&0&$\cdots$&0&1&-2&1&$\cdots$&0&$Q_1$\\  \hline
     $p_2$&0&1&0&$\cdots$&0&0&2&-4&$\cdots$&0&$Q_2$\\  \hline
     $p_3$&0&0&1&$\cdots$&0&0&0&3&$\cdots$&0&$Q_3$\\  \hline
    $\vdots$&$\vdots$&$\vdots$&$\vdots$&$\ddots$&$\vdots$&$\vdots$&$\vdots$&$\vdots$&$\ddots$&$\vdots$&$\vdots$\\  \hline
    $p_{d}$& 0&0&0&$\cdots$&1&0&0&0&$\cdots$&d&$Q_d$\\
    \hline
   c& 0&0&0&$\cdots$&0&$c_{d+1}$&$c_{d+2}$&$c_{d+3}$&$\cdots$&$c_{2d}$&$Q$\\
    \hline
    \end{tabular}
  \end{center}
\end{table}
In Table III, there are $c_{d+i}=(i-2)f(i-2)+if(i)-2(i-1)f(i-1)$ and $Q_i=b_{i(d+i)}q_i+b_{i(d+i+1)}q_{i+1}+b_{i(d+i+2)}q_{i+2},$ where $b_{i(d+i)}=i$, $b_{i(d+i+1)}=-2i$, and $b_{i(d+i+2)}=i$. Here, $b_{ij}$ corresponds to the $i$-th element of the vector $\textbf{b}_j$. With these relations, we then obtain $p_i=Q_i=i(\abs{c_i}^2-\abs{c_{i+1}}^2)$.  Since there are $c_{d+i}=(i-2)f(i-2)+if(i)-2(i-1)f(i-1)\geq0$ for all $i=1,\cdots,d$, we then obtain the optimal solution, i.e.,
\begin{eqnarray}
\overline{C}_{\max}(\varphi_\mu^\text{m})=Q=\textbf{c}^T\textbf{p}=\sum_{i=1}^di(\abs{c_i}^2-\abs{c_{i+1}}^2)f(i). \label{value1}
\end{eqnarray}


\begin{thebibliography}{99}
\bibitem{note} We should note that the coherence rank defined in \cite{Winter}, $C_R(\varphi)=R$, does not fulfill (C1). We here defined it as $C_R(\varphi)=R-1$ to avoid this.
\bibitem{note2} To avoid the negativity, we may transform the  $p_i$  corresponding to $\{i|\min_iq_j/b_{ij}~\text{with}~b_{ij}>0\}$ from a free variable into a basic variable (See Ref. \cite{Luenberger} for details).
\end{thebibliography}
\end{document}